\documentclass[%
 reprint,
 amsmath,amssymb,
 aps,
]{revtex4-2}

\usepackage{graphicx}% Include figure files
\usepackage{dcolumn}% Align table columns on decimal point
\usepackage{bm}% bold math
\usepackage{textalpha}
\usepackage{comment}
\usepackage{color}

\begin{document}
\preprint{APS/123-QED}

%\title{Linear viscoelastic spectra of soft particulate gels: master curve and physical origin of the fractal constitutive behavior}% Force line breaks with \\

\title{Microscopic interactions and emerging elasticity in model soft particulate gels}

\author{Minaspi Bantawa$^1$, Wayan A. Fontaine-Seiler$^1$, Peter D. Olmsted$^1$, and Emanuela Del Gado$^1$}

\address{$^1$Department  of  Physics,  Institute  for Soft  Matter  Synthesis and Metrology,
Georgetown  University,  37th and O Streets,  N.W., Washington,  D.C. 20057,  USA}
%\ead{mb1891@georgetown.edu and ed610@georgetown.edu}
\vspace{10pt}
%\begin{indented}
%\item[]July 2021
%\end{indented}

\begin{abstract}
We discuss a class of models for particulate gels in which the particle contacts are described by an effective interaction combining a two-body attraction and  a three-body angular repulsion. Using molecular dynamics, we show how varying the model parameters allows us to sample, for a given gelation protocol, a variety of gel morphologies. For a specific set of the model parameters, we identify the local elastic structures that get interlocked in the gel network. Using the analytical expression of their elastic energy from the microscopic interactions, we can estimate their contribution to the emergent elasticity of the gel and gain new insight into its origin. 
\end{abstract}

%\vspace{2pc}
%\noindent{\it Keywords}: Particulate gels, Short-range attraction, Bending rigidity, Coordination number, Network topology, Elastic modulus
\maketitle
\section{Introduction}
Soft particulate gels can form in a wide range of colloidal suspensions of particles, particle agglomerates or droplets \cite{DelGadoBookChapter2016, Trappe2001,Dibble2008,Campbell2005,Stradner2004,laurati2009, Ako2010,Tanaka2005Protein,DICKINSON2012,Peng2019, Helgeson2014}. The combination of attractive inter-particle interactions, that drive the particle aggregation, with increasing cooperative dynamics, that lead to kinetic arrest, results into the self-assembly of an interconnected space-spanning network structure, which can be very soft but ultimately has solid-like elastic properties \cite{delgado2003unifying,delgado2004slow}. These gels have relevance to a wide variety of industries and technologies, ranging from food or personal care products to drug delivery or other biomedical applications \cite{Storm2005,Tanaka2005Protein,DICKINSON2012,Helgeson2014}.

In spite of their relevance, the mechanics and rheology of these materials is still hard to predict, because the extreme variability of microstructural characteristics makes it difficult to identify common underlying mechanisms. Extensive work has been devoted to characterizing gel morphologies and processes that can initiate gel formation, from irreversible and diffusion limited colloidal aggregation with fractal-like microstructures \cite{Shih1990,Wu2013}, to phase separation or microphases \cite{Tanaka2005,Tanaka2005Protein,Lu2008,Helgeson2014}. Extensive work exists also on the cooperative and glassy microscopic dynamics associated to gel formation \cite{delgado2004slow,DelGado2010,laurati2009,Cipelletti2005,Doorn2017}, elucidating how this is a common trait for these systems, quite independently from the specific mechanism that initiates gelation. Different gelation processes, however, can change the amount of stress heterogeneities frozen-in in the gel structure upon solidification. The possibility to relax those stresses through the elasticity stored in the gel network can make the glassy dynamics of soft gels quite distinct, with faster-then-exponential relaxations and intermittent spatio-temporal correlations \cite{Cipelletti2001,Duri2005, Bouzid2017}. Extensive experimental work and more recent simulation studies have addressed the connections between microstructure and rheological behavior \cite{laurati2009,Colombo:2014JOR,zia2014,moghimi2017}, also considering hydrodynamic interactions through the solvent in which the particles are embedded \cite{Varga2015JOR,Wang2019,Arman-jamali2018}. The outstanding questions %go from
at this point include identifying a possible common origin of the solid-like mechanical response across materials apparently very different and understanding the emergence of rigidity and of stress localization \cite{delgado2021-ency,Johnson2021}. 

Here we are interested especially in these last two issues and in the contributions that numerical simulations of simple but judiciously designed statistical mechanical models can give \cite{Bouzid:2018BookChap}. To this aim, we have designed a model for soft particulate gels that can cover a range of interparticle interactions and can be studied through coarse-grained molecular dynamics simulations, where different gelation processes can be mimicked. This approach seems promising to explore how different gel morphologies can emerge due to the interplay between microscopic interactions and kinetic processes. It has already proven successful in disentangling the role of structural and stress heterogeneities in the microscopic glassy dynamics \cite{Colombo2013,Colombo:2014SM, Bouzid2017} and linear or non-linear rheology \cite{Colombo:2014JOR,Bouzid:2018LAN,Feng:2018PNAS,Vereroudakis2020} of colloidal gels, providing useful insight into experimental observations.  

In this paper, we study how the specific ingredients of the microscopic model translate into the elasticity of different parts of the gel structure, and how varying the model parameters allows us to obtain a range of gel microstructures relevant to real materials. The paper is organized as follows. Section \ref{section2} introduces the model, and the simulation method used for the gel preparation. Section \ref{parameters} describes the changes in gel microstructures by varying the model parameters. In Section \ref{elasticity}, we discuss the elasticity of different structural elements of the gel starting from the microscopic interactions, and analyze their contribution to the gel network elasticity. Finally, section \ref{conclusions} contains conclusions and an outlook.

%%%%%%%%%%%%%%%%%%%%%%%%%%%%%%%%%%%%%%%%%%%%%%%%%%%%%%%%%%%%%%%%%%%%%%%%%%
\section{Model choice and numerical simulations}\label{section2}

The net attractive interactions between particles or particle agglomerates in solution usually originate from surface forces such as those described by the Derjaguin, Landau, Vervey, and Overbeek (DLVO) theory \cite{DLVO1948,Israelachvili} with van der Waals forces being the main source of attraction \cite{Israelachvili}. In some cases they can also originate from an entropically favored depletion, from the interparticle gaps, of % from the interparticle spacings, of 
small non-adsorbing polymers added to the the suspension, as described by the Asakura-Oosawa theory \cite{AO1954}. While these theories usually capture %well 
the energy scales and interaction ranges measured in experiments, they are intrinsically mean field in nature and do not include aspects of the particle surfaces that can become important once the particles are in close contact, a regime important for gel formation. 

\begin{figure}[htp]
\centering
\includegraphics[scale=0.145]{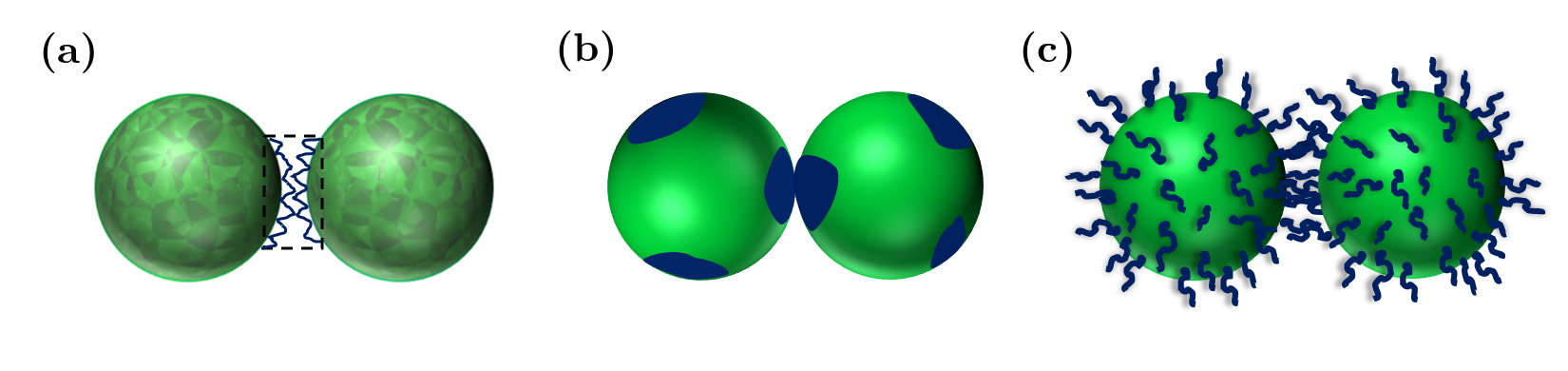}
\caption{Different examples of particle contacts that can give rise to bending rigidity in colloidal gels: (a) Particles with rough surfaces get interlocked. (b) %Patchy particles stick with each other only at their patches. 
Heterogeneous surface patches lead two particles to stick at specific sites. (c) Relative rotation of particles with surfaces grafted by polymer chains in close contact can be hindered by the chain overlap.}
\label{schematic}
\end{figure}

In many real materials, the particle surfaces are not smooth or homogeneous (see cartoons in Fig. \ref{schematic}). When particles with surface irregularities or aggregates of particles come in contact, the surface roughness or a local deformation can lead to the interlocking of the surfaces (see Fig. \ref{schematic} (a)). In some cases, surface irregularities are present in the form of surfaces patches that behave as specific binding sites (Fig. \ref{schematic} (b)). In other cases, colloidal particles are sterically stabilized by adsorbed or grafted polymer chains (Fig. \ref{schematic} (c)), which can hinder relative sliding or rotation. The same can happen for compact aggregates of irregular shape or fractal-like flocs with reduced connectivity.  All these different cases can lead to an effective bending rigidity of parts of the gel structure, as the gel self-assembles. There is evidence of these phenomena from confocal microscopy images in experiments, showing that local coordination of particulate gels can be limited to 2-4 contacts even when there is not a clear fractal characteristics of the gel structure. Optical tweezers experiments have proven that strands of aggregated colloidal particles can sustain finite torques, and it has been recently shown that the mechanical contacts between colloidal particles can be solid-on-solid contacts, which stiffen over time \cite{Dibble2008,Campbell2005,Ohtsuka2008,Wu2020,Whitaker2019,Bonacci2020}. 

%%%%%%%%%%%%%%%%%%%%%%%%%%%%%%%%%%%%%%%%%%%%%%%%%%%%%%%%%%%%%%%%%%%%%%%
%%%%%%%%%%%%%%%%%%%%%%%%%%%%%%%%%%%%%%%%%%%%%%%%%%%%%%%%%%%%%%%%%%%%%%

A physical model for computer simulations that has the goal to gain new insight into the gel mechanics and the underlying microscopic mechanisms should include these possible effects. However, to be able to effectively perform large scale simulations and extended spatio-temporal analysis of microscopic processes, one would like to avoid a fully atomistic description of the particle contacts. With this in mind, we have introduced a class of microscopic models for soft particulate gels that features a short-range attraction, similar to the one predicted by several theories of colloidal interactions, and an additional term that depends on the angle between particle bonds, to include the energy costs associated with the constraints of the particle relative motion imposed by the nature of the surface contacts \cite{DelGado2010,Colombo2013,Colombo:2014SM,delgado2005}. In previous studies, we have shown how this approach can help to understand the microscopic origin of the complex relaxation dynamics \cite{DelGado2010,Colombo2013,Colombo:2014SM}, aging \cite{Bouzid2017} and mechanical response \cite{Colombo:2014JOR,Bouzid:2018LAN,Feng:2018PNAS,BouzidJOR2018,Vereroudakis2020} in colloidal gel networks. Theses studies have demonstrated that the dynamical and mechanical properties in these materials emerge from mesoscale structural characteristics of the gel networks, providing an explanation to the observation of common traits found across different materials. Recent numerical studies from different groups have also confirmed that including similar constraints, in addition to the attraction strength and range that can be justified with existing theories of colloidal interactions, is essential to properly reproduce the characteristics of the mechanics of soft particulate gels \cite{Saw2009,CGARLEA2017,Bianchi2008,Wang2019,Foffi2005,Kern2003,Immink2020,HongT2020,Hamed2020}. 
%%%%%%%%%%%%%%%%%%%%%%%%%%%%%%%%%%%%%%%%%%%%%%%%%%%%%%%%%%%%%%%%%%%%%%%

\begin{figure*}[htp]
\centering
\includegraphics[scale=0.2]{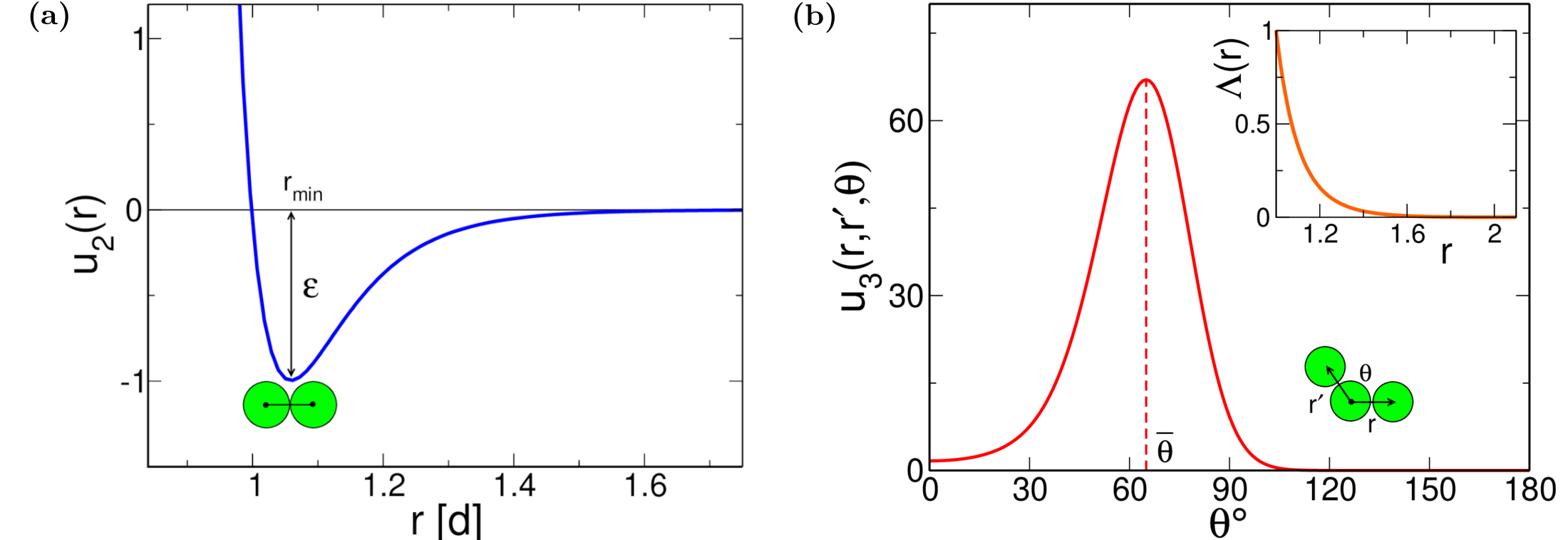}
\caption{(a) Two-body potential $u_2$ as a function of distance between two particles, $r$ in units of a particle diameter $d$, and (b) Main: Three-body interaction $u_3$ for  $r = r^\prime = r_{min}$ (both bonds at the minimum of the attractive well) and $\overline{\theta}=65^{\circ}$ as a function of angle between the neighboring bonds, $\theta$. Other parameters are $B = 67.27$, and $w = 0.30$. Inset: Radial modulation $\Lambda(r)$ as a function of distance.}
\label{potentials}
\end{figure*} 
%%%%%%%%%%%%%%%%%%%%%%%%%%%%%%%%%%%%%%%%%%%%%%%%%%%%%%%%%%%%%%%%%%%%%%

\subsection{Numerical model}\label{model}
The model consists of $N$ identical particles each of diameter $d$, and described as point-like, whose coordinates are $\{\textbf{r}_i\}$, with $i = 1,...,N$. They interact via a potential energy \cite{Colombo2013,Colombo:2014JOR,Bouzid2017}:
%%%%%%%%%%%%%%%%%%%%%%%%%%%%%%%%%%%%%%%%%%%%%%%%%%%%%%%%%%%%%%%%%%%%%%%
\begin{equation}\label{Potential}
U({\textbf{r}}_1,..,{\textbf{r}}_N) = \varepsilon \Bigg [{\sum_{i>j}}u_2 \Big (\frac{\textbf{r}_{ij}}{d} \Big ) + {\sum\limits_i}{\sum\limits_{j>k}^{j,k \neq i}} u_3 \Big (\frac{\textbf{r}_{ij}}{d},\frac{\textbf{r}_{ik}}{d} \Big ) \Bigg ]
\end{equation}
%%%%%%%%%%%%%%%%%%%%%%%%%%%%%%%%%%%%%%%%%%%%%%%%%%%%%%%%%%%%%%%%%%%%%%%%
where $\textbf{r}_{ij} =  \textbf{r}_j - \textbf{r}_i$ is the vector separation between two particles $i$ and $j$, $\varepsilon$ is the depth of the attractive well in $u_2$, used as unit energy in the simulations. For colloidal suspensions, the value of $d$ is generally in the range $10 - 100$ nm and $\varepsilon \simeq 1 - 100$ $k_BT$ \cite{Dibble2008,Trappe2001}, where $k_B$ and $T$ indicate the Boltzmann constant and the room temperature.\\
\indent The two-body term $u_2$ in Eq. (\ref{Potential}) is a Lennard-Jones (LJ) like potential, and is a combination of a repulsive core and a narrow attractive well. For particles separated by a distance $r$ (here and in the following, distance is expressed in units of $d$), it is written in the form:
%%%%%%%%%%%%%%%%%%%%%%%%%%%%%%%%%%%%%%%%%%%%%%%%%%%%%%%%%%%%%%%%%%%%%%%
\indent \begin{equation}\label{Two-body term}
u_2(r) = 23 \Big ( \frac{1}{r^{18}} - \frac{1}{r^{16}}  \Big ) 
\end{equation}
%%%%%%%%%%%%%%%%%%%%%%%%%%%%%%%%%%%%%%%%%%%%%%%%%%%%%%%%%%%%%%%%%%%%%%%%
for computational convenience. The values of the exponents $18-16$ in this generalized LJ form have been chosen to produce a short range attractive well (less than 1.5 particle diameters, with a minimum $r_{min}\sim1.06d$), which is plotted in Fig. \ref{potentials} (a).    

\indent The three-body term $u_3$ in Eq. (\ref{Potential}) represents the angular repulsion (directional interaction) which constraints the possible configurations of particles bonded to a central one, providing bending rigidity to inter-particle bonds ${\textbf{r}}$ and $\textbf{r}^{\prime}$ departing from the same particle (see Fig. \ref{potentials} (b)). The functional form of this term has been implemented, again considering computational efficiency, as:
%%%%%%%%%%%%%%%%%%%%%%%%%%%%%%%%%%%%%%%%%%%%%%%%%%%%%%%%%%%%%%%%%%%%%%%%%
\indent \begin{equation}\label{Three-body term}
 u_3(\textbf{r},\textbf{r}^\prime) =  B\Lambda(r)\Lambda(r^\prime)\exp\Bigg [-\bigg(\frac{\textbf{r} \cdot \textbf{r}^\prime} {rr^\prime}-\cos\overline{\theta}\bigg)^2 \bigg / w^2 \Bigg]
\end{equation}
%%%%%%%%%%%%%%%%%%%%%%%%%%%%%%%%%%%%%%%%%%%%%%%%%%%%%%%%%%%%%%%%%%%%%%%%%
where $B$, $\overline{\theta}$ and $w$ are dimensionless parameters. The parameter $B$ represents the strength of this interaction term. With this specific form we aim at maintaining the possibility of a range of allowed configurations, without imposing only one specific angle. The radial modulation function $\Lambda(r)$ is plotted in Fig. \ref{potentials} (b) (inset) and decays smoothly as,
%%%%%%%%%%%%%%%%%%%%%%%%%%%%%%%%%%%%%%%%%%%%%%%%%%%%%%%%%%%%%%%%%%%%%
\begin{equation}\label{lambda}
\Lambda(r)= r^{-10}\left[1-(r/2)^{10}\right]^2 {\mathcal{H}}(2-r)
\end{equation}
%%%%%%%%%%%%%%%%%%%%%%%%%%%%%%%%%%%%%%%%%%%%%%%%%%%%%%%%%%%%%%%%%%%%%%
where ${\mathcal{H}}$ is the Heaviside function.
The function $\Lambda(r)$ vanishes at a distance $2d$, but, combined with $u_2$, it effectively cancels out the attraction $u_2(r)$ producing a repulsion that vanishes at a distance $\simeq d$ or leaves an attractive well basically identical to $u_2$, depending on the angle in the exponential term of Eq.~\ref{Three-body term}. That is, the additional repulsion introduced by $\Lambda(r)$ is able to cancel out the attractive well (in the angular range needed), while retaining the continuity of energy and force required by the molecular dynamics method. 

In Fig. \ref{potentials} (b), the three-body potential $u_3$ is plotted as a function of the bending angle $\theta$ by fixing the bond distances at the minimum of the potential well, i.e. $r=r^\prime = r_{min}$. The height of the peak is determined by the parameter $B$ and represents the strength of angular repulsion. The peak location is determined by the angle $\bar{\theta}$ and its width is determined by $w$.  

%%%%%%%%%%%%%%%%%%%%%%%%%%%%%%%%%%%%%%%%%%%%%%%%%%%%%%%%%%%%%%%%%%%%%%
\begin{figure}[htp]
\centering
\includegraphics[scale=0.11]{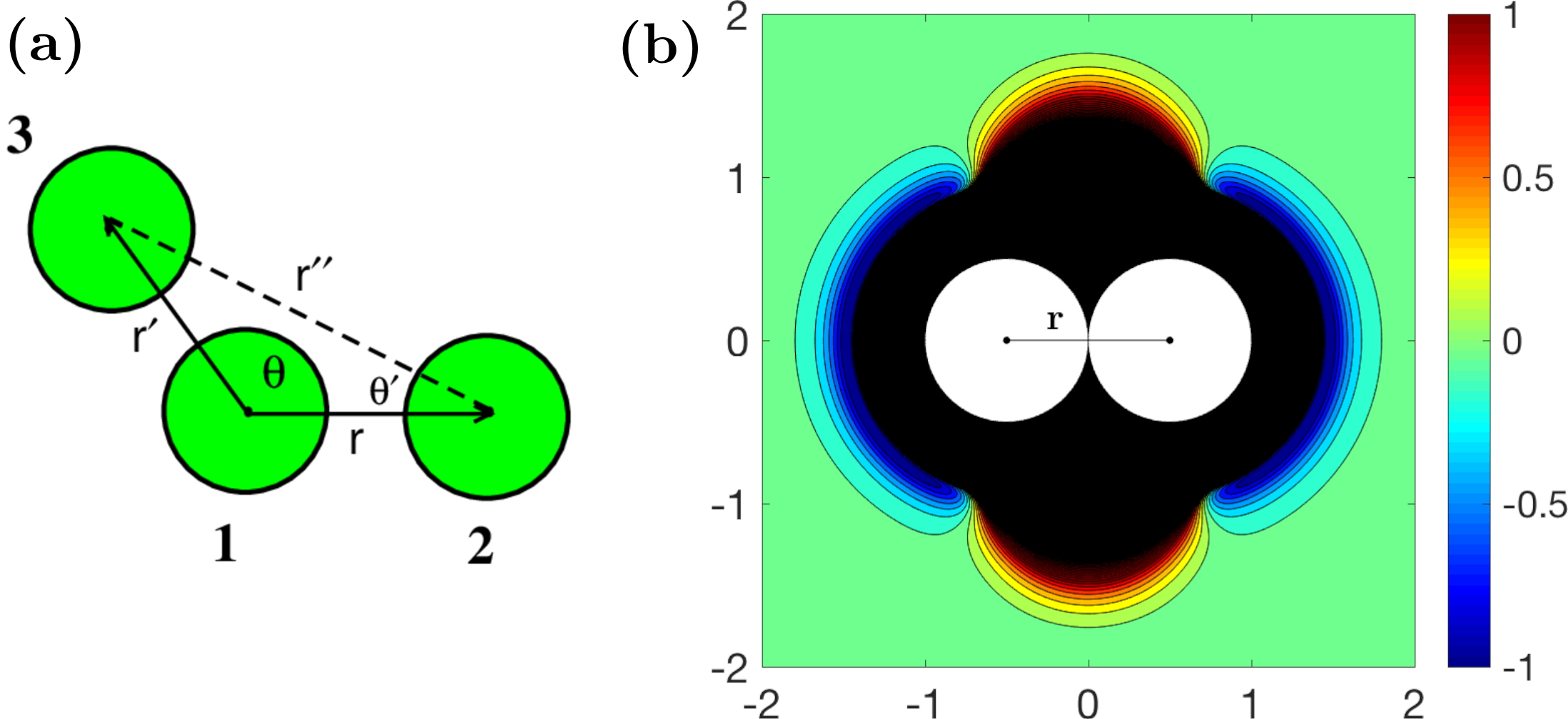}
\caption{(a) Illustration of the interactions for a particle \textbf{3} approaching two particles \textbf{1} and \textbf{2} previously bonded. (b) Contour plot of the potential energy experienced by an incoming particle when it approaches an existing bond at different distances and angles. The color is blue when the potential energy is $-1$ and is red for $+1$ or higher.}\label{fig2}
\end{figure}
%%%%%%%%%%%%%%%%%%%%%%%%%%%%%%%%%%%%%%%%%%%%%%%%%%%%%%%%%%%%%%%%%%%%%%

In order to illustrate how the combination of $u_2$ and $u_3$ works, let's consider the case where all parameters have been fixed as in Fig.\ref{potentials}. In Fig. \ref{fig2} (a), we consider a bond vector $\textbf{r}$ formed by two particles, \textbf{1} and \textbf{2}, which are within distances corresponding to the attractive well. A third particle, \textbf{3}, approaches particle \textbf{1} from a direction such that the vector distance ${\textbf{r}}^\prime$ from particle \textbf{1} to \textbf{3} forms an angle $\theta$ with $\textbf{r}$. If the incoming particle is within the range of interaction with particle \textbf{1}, it experiences an attractive interaction given by the two-body term, $u_2(r)$, and an angular repulsion given by the three-body interaction $u_3(r,r^\prime,\theta)$ that involves both bonds ${\textbf{r}}^\prime$ and $\textbf{r}$. Due to the angular modulation of $u_3(r,r^\prime,\theta)$, the particle \textbf{3} will be bonded to \textbf{1} or \textbf{2} only for certain range of angles $\theta$ where the net interaction is attractive. The range of the angles for the net attraction is controlled by the parameters $B$ and $\overline{\theta}$ in Eq. \ref{Three-body term}. For the choice of $B=67.27$ and $\bar{\theta}=65^\circ$ used in the figure, \textbf{3} can not be simultaneously bonded to \textbf{1} and \textbf{2}. All angles $\theta$ smaller than $\overline{\theta}$ are disfavored %completely forbidden 
when ${\textbf{r}}^\prime$ is within the range of interaction with \textbf{1} because, for smaller $\theta$, particle \textbf{3} will be sufficiently close to particle \textbf{2} to experience the short-range repulsion from $u_2(r^{\prime \prime})$. On the other hand, if particle \textbf{3} approaches \textbf{1} at smaller angles, but at distances larger than the range of interaction, it will be close enough to particle \textbf{2} to remain bonded there. In this case, the interactions between \textbf{1} and \textbf{3} becomes negligible. To summarize, when the particle \textbf{3} enters %at different possible range of angles and distances into 
the region of the bond between \textbf{1} and \textbf{2}, the total potential felt $u_{total}$ is obtained by summing up all two- and three-body interactions:
 %%%%%%%%%%%%%%%%%%%%%%%%%%%%%%%%%%%%%%%%%%%%%%%%%%%%%%%%%%%%%%%%%%%%%%%
 \begin{align}\label{utotal1}
 u_\mathrm{total}&=
 \begin{aligned}[t]
& u_2(r^\prime)+u_3(r,r^\prime,\theta)+u_2(r^{\prime \prime})\\
& +u_3(r,r^{\prime \prime},\theta^\prime)+u_3(r^\prime,r^{\prime \prime},\pi-\theta-\theta^\prime)\\
 \end{aligned}
 \end{align}
%%%%%%%%%%%%%%%%%%%%%%%%%%%%%%%%%%%%%%%%%%%%%%%%%%%%%%%%%%%%%%%%%%%%%%%%%
We can rewrite this potential energy as $u_\mathrm{total} = \sum u_2(r,r^\prime, \theta)+\sum u_3(r,r^\prime, \theta)$, considering that $r^{\prime \prime}=\sqrt{r^2+{r^\prime}^2-2rr^\prime \cos{\theta}}$ and $\theta^\prime=\cos^{-1}[(r-r^\prime\cos{\theta})/(\sqrt{r^2+{r^\prime}^2-2rr^\prime\cos{\theta}})]$.  

In Fig. \ref{fig2} (b), we fix the distance between particles \textbf{1} and \textbf{2} at $r=r_{min}$ and show the contour plot of $u_\mathrm{total} (r_{min}, r\prime,\theta)$, using $x=r^\prime \cos\theta$ and $y=r^\prime \sin\theta$ for the same choice of $B$, $\bar{\theta}$ and $w$ . The colors highlight the modulation from attraction to repulsion depending on distance and angles, as particle \textbf{3} approaches \textbf{1} and \textbf{2}. It is symmetric with respect to the original bond formed by particles \textbf{1} and \textbf{2}) i.e., particle \textbf{3} can equally stick to any of the other two particles depending on the direction as dictated by the region in blue, where the potential is attractive. 

In section \ref{parameters} below, we will discuss further how changing the parameters $B$, $\bar{\theta}$ and $w$ can change these potential energy profiles and hence modify the type of structures obtained in the simulations.

%%%%%%%%%%%%%%%%%%%%%%%%%%%%%%%%%%%%%%%%%%%%%%%%%%%%%%%%%%%%%%%%%%%%%%%
\begin{figure*}[btp]
\centering
\includegraphics[scale=0.1]{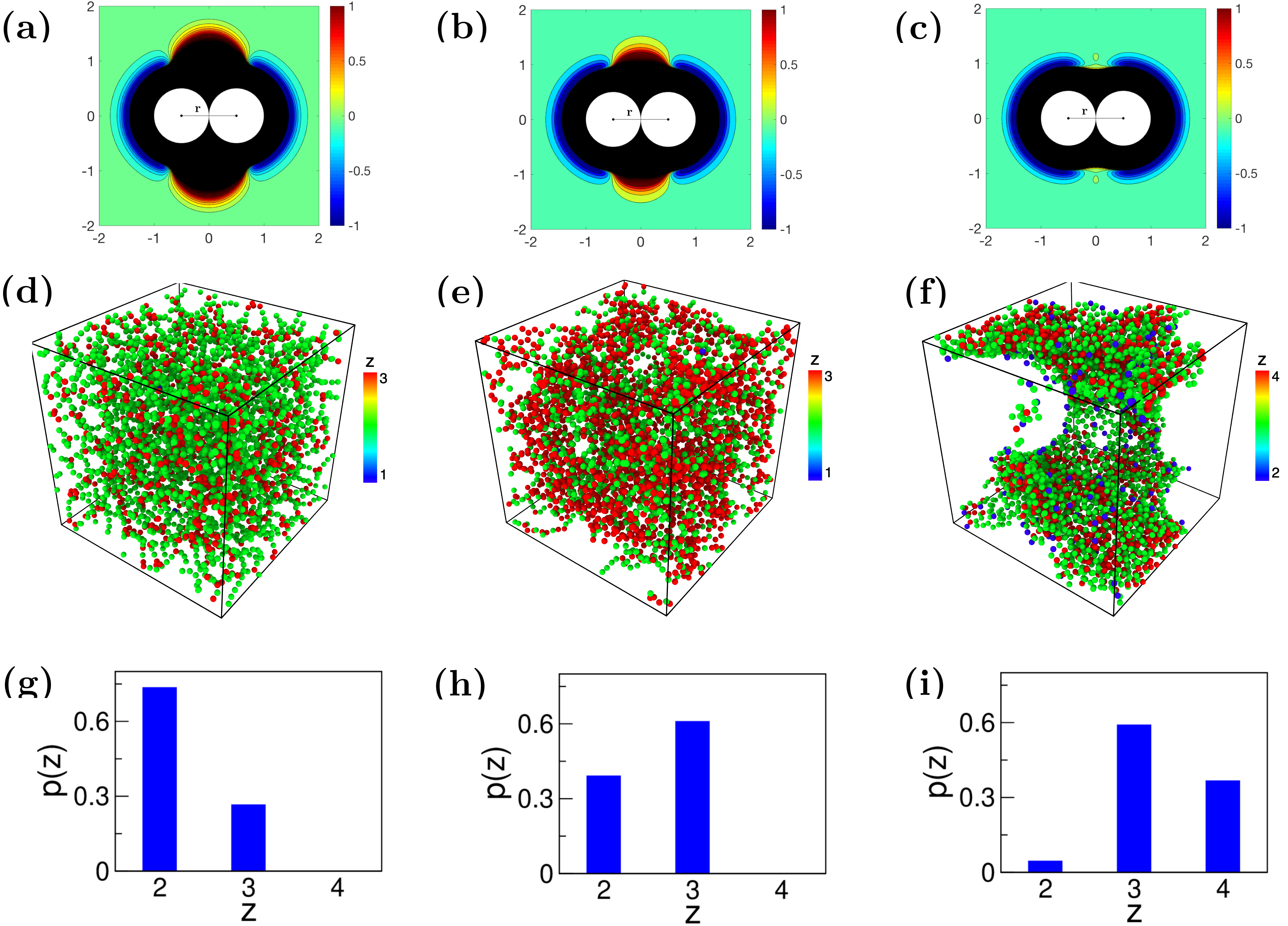}
\caption{Potential energy contour plots for $B=67.27$, $10$ and $1$ in (a)-(c), Simulation snapshots of gels, each at a volume fraction $\phi=7.5\%$ corresponding to above values of $B$ in (d)-(f) where the color code represents the local coordination number $z$, and the distribution of $z$ of the final gel structures in (g)-(i). The distributions are obtained by averaging over $5$ independently generated structures and the error bars (obtained from sample-to-sample fluctuations) are smaller than the bar thickness.}
\label{changeB}
\end{figure*}
%%%%%%%%%%%%%%%%%%%%%%%%%%%%%%%%%%%%%%%%%%%%%%%%%%%%%%%%%%%%%%%%%%%%%%

\subsection{Numerical simulations of gel preparation}\label{preparation}

In the Molecular Dynamics (MD) simulations, we use $N = 16384$ particles in a cubic box of size $L$ and number density $N/L^3$. If we consider each particle to be a sphere of diameter $d$, we can define an approximate solid volume fraction $\phi =\frac{N \pi d^3/6}{(Ld)^3}$. Here we discuss gels at volume fractions $\phi=5\%, 7.5\%, 10\%, 15\%$. We use periodic boundary conditions and solve the equations of motion with the interactions described in section \ref{model} and a time step $\delta t = 0.005\tau_0$ ($\tau_0 = \sqrt{md^2/\epsilon}$ is the usual MD time unit).

The initial gel configurations are prepared in two steps. The first step involves cooling of a system of particles previously equilibrated at a reduced temperature $k_{B}T/\varepsilon = 0.5$ in a gas phase to a temperature $k_{B}T/\varepsilon = 0.05$, low enough for the particles to aggregate and form a gel network. We use NVT equilibrium MD simulations, with a Nosé-Hoover (NH) thermostat and a cooling rate of $\Gamma = \frac{\Delta T}{\Delta t} \approx 10^{-4} \varepsilon /k_B\tau_0$. We have verified that, in this temperature regime, the microstates obtained do not significantly depend on the dynamics used, for the cooling rate we consider here. Using the simple NVT MD in this context is therefore more convenient, because it reduces the simulation time with respect to a more physical Langevin dynamics. Then, we let the system further equilibrate at $k_{B}T/\varepsilon = 0.05$ with the NH thermostat for another $10^6$ MD steps until all the structural quantities reach a steady state and any further aging of the gel takes place over much longer time scales. 

In the second part of the gel preparation, we use instead a damped dynamics to drive the gel network, formed at finite temperature, to a local minimum that more likely corresponds to a mechanically stable configuration. This is done by draining the kinetic energy to $\sim 10^{-10}$ of its initial value with an overdamped dissipative dynamics:

\indent \begin{equation} \label{Damped dynamics}
m\frac{d^2\textbf{r}_i}{dt^2} = -\zeta\frac{d\textbf{r}_i}{dt} - {\bf{\nabla}}_{\textbf{r}_i}U
\end{equation}
where $m$ is the mass of the particle and $\zeta$ is the drag coefficient for the solvent \cite{Colombo:2014JOR,Bouzid:2018LAN}. We used $m/\zeta$ = 1 in all the simulations discussed here. This part could be, in principle, also achieved by minimizing the total energy of the system with a conjugate gradient algorithm. However, we have previously found that using a damped dynamics is actually more efficient with very soft gels, as discussed in \cite{Colombo:2014SM,Colombo:2014JOR,Bouzid:2018LAN}. For simplicity, we have therefore used this damped dynamics for the energy minimization in all simulations.

Starting from different initial positions and velocities at high temperature, we generate $5$ statistically independent samples with this same two-step procedure and use them to perform ensemble averages of all quantities studied here. For each samples, we study its steady-state dynamical evolution using the Langevin dynamics which includes the effect of the solvent and of thermal fluctuations, given by the equation:
\begin{equation} 
\label{Langevin dynamics}
m\frac{d^2\textbf{r}_i}{dt^2} = -\zeta\frac{d\textbf{r}_i}{dt} -
%{\bf{\bigtriangledown}}_{\textbf{r}_i}U+\textrm{F}_{\textbf{r}_i}(t)
\nabla_{\textbf{r}_i}U+\textrm{F}_{\textbf{r}_i}(t)
\end{equation}
where $m/\zeta$ = 1. $\textrm{F}_{\textbf{r}_i}$ is a random force that introduces the thermal fluctuations and sets the temperature $T$ through the equation:
$\langle F_r^i(t) F_r^j(t^\prime)\rangle=2\zeta k_B T\delta_{ij} \delta(t-t^\prime)$. To sample the bond angle distributions in gel networks, we run these simulations at $k_{B}T/\varepsilon = 10^{-4}$ for $3\cdot10^{6}$ MD steps. The distribution of coordination number $z$ and contour length $l_C$ of the gel strands are computed for $5$ independently generated configurations at their energy minima, since there is no change during the dynamics over the simulation window considered here. The distribution of coordination numbers $z$ is defined by $p(z)=N_z/N$,  where $N_z$ is the number of particles that have specific coordination number $z$ and $N$ is the total number of particles. The elastic moduli $G_0$ of gel networks at different volume fractions are computed for $3$ independently generated configurations also at their energy minima by applying an oscillatory shear strain and measuring the stress response as described in  \cite{Bouzid:2018BookChap, BouzidJOR2018}.

We use a slightly different procedure to prepare an isolated chain or strand, since in this case the particles are placed in a chain to start with. Then, we minimize the configuration energy with the damped dynamics (Eq. \ref{Damped dynamics}) used for the gels, and then use the Langevin dynamics (Eq. \ref{Langevin dynamics}) to sample the configurations at fixed $k_BT/\varepsilon$, for $2\cdot10^8$ MD steps, and obtain the bond angle distributions and the persistence length. In a $3$ particle strand, the bond angle distributions are computed at temperatures $k_BT/\varepsilon=10^{-5}, 10^{-4}, 10^{-3}, 10^{-2}$. The persistence length is determined from a $50$-particle strand, by computing the correlation in the angles of successive bonds along the strand \cite{Colombo:2014SM,Feng:2018PNAS}. All the simulations have been performed using a modified version of LAMMPS \cite{PLIMPTON:1995JCP} that includes the interactions discussed in section \ref{model}. All simulations performed are summarized in Table \ref{table1}.
\begin{table}[h!]
\centering
\caption{{Simulations performed}}
\label{table1}
\begin{tabular}{|c|c|c|}
\hline
{Structure} & {Preparation} & {Data production}\\
\hline

{Network} & {NH+ Dissipative} & {Langevin}\\
{} & {} & {($3\cdot 10^6$ MD steps)}\\
\hline
{Isolated strands} & {Dissipative} & {Langevin)}\\
{} & {} & {($2\cdot10^8$ MD steps)}\\
\hline
\end{tabular}
\end{table}

\section{Varying the model parameters}\label{parameters}

%To elucidate the role of the $u_{3}$ term in the interactions, we %keep all parameters of $u_{2}$ fixed and change the parameters in $u_{3}$. 
Now we discuss the implications of the parameters choices for $B$, $w$ and $\overline{\theta}$ in $u_{3}$ in terms of how they modify the potential energy profiles shown in Fig.\ref{fig2}(b) and the gel structures obtained in the simulations. Figs. \ref{changeB}(a)-(c) show contour plots for $u_\mathrm{total}$ and snapshots of gels, all at a volume fraction $\phi=7.5\%$ and obtained with the procedure described in section \ref{preparation}, with decreasing $B$ (left to right) while keeping $\overline{\theta}=65^{\circ}$ and $w = 0.3$. The plots (a), (d) and (g) correspond to the same set of parameters discussed in section \ref{model}. With decreasing $B$, the repulsive barrier shown in Fig. \ref{fig3}(b) becomes weaker while the region in blue, corresponding to the attractive well in the contour plots, becomes wider, i.e. decreasing $B$ also changes the angular modulation of the $u_{3}$ term. A consequence is that the resulting gel structures will change from a thin, space filling network with coordination number $z$ mostly $2$ or $3$ as also seen in \cite{Dinsmore2002,Campbell2005,Dibble2008,Ohtsuka2008, Eberle2011}, to locally compact aggregates with higher $z$ and large pores typical of phase separation as found in \cite{Tanaka2005,Tanaka2005Protein,Lu2008,Gibaud2009,Gibaud2013}. The distribution $p(z)$ of the coordination numbers $z$, for each of the structures are plotted in Figs. \ref{changeB}(g)-(i).

To further elucidate the role of the angle parameter $\bar{\theta}$ and its implications for the possible gel microstructures, we compare simulations performed, for the same preparation protocol, with $\bar{\theta} = 65^\circ$ and $\bar{\theta}=75^\circ$, while the others parameters are kept as in Fig.~\ref{fig2}. 
%%%%%%%%%%%%%%%%%%%%%%%%%%%%%%%%%%%%%%%%%%%%%%%%%%%%%%%%%%%%%%%%%%%%%%%
\begin{figure}[htp]
\centering
\includegraphics[scale=0.09]{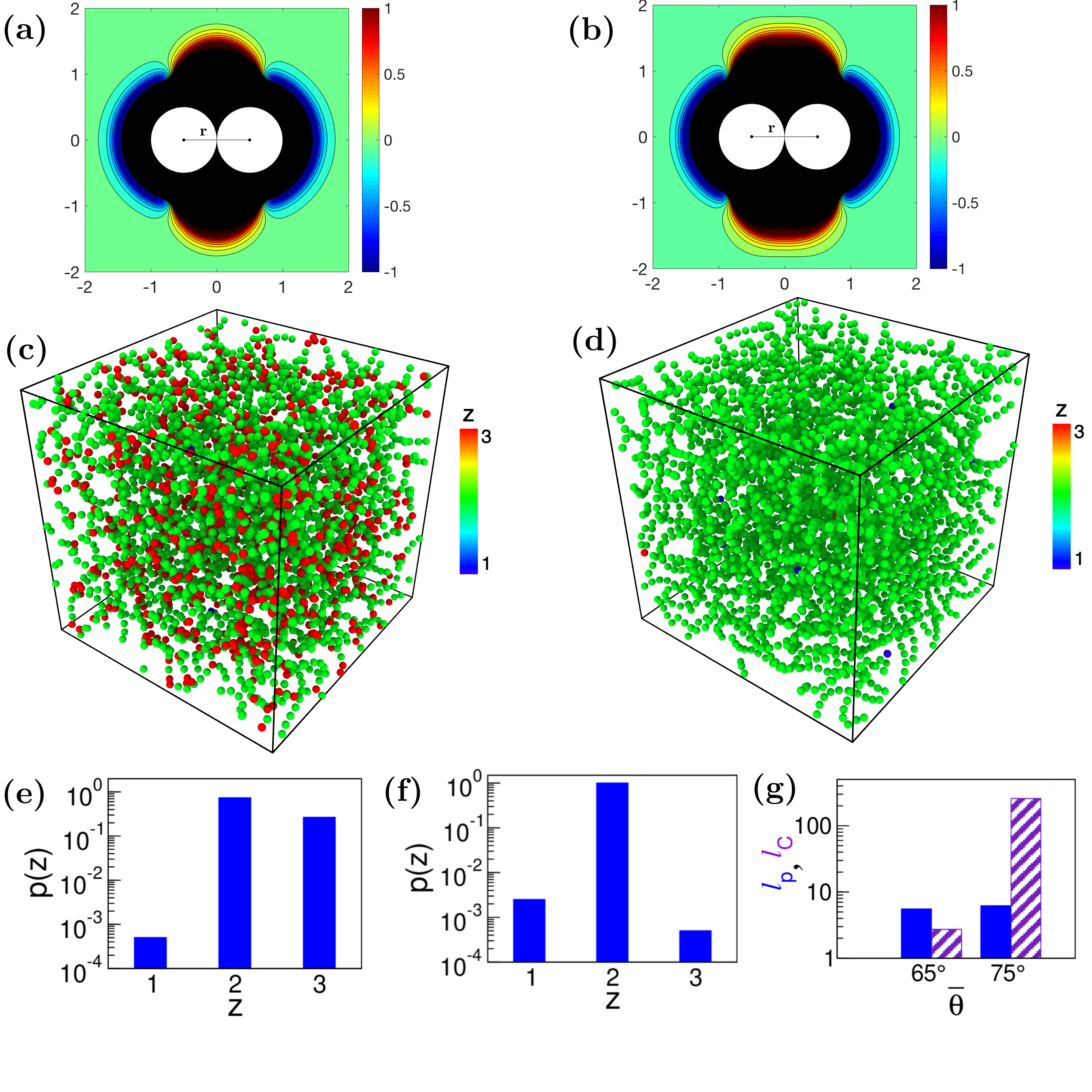}
\caption{Potential energy contour plots for $\bar{\theta} = 65^\circ$ and $75^\circ$ in (a) and (b), corresponding simulation snapshots of gels, each at a volume fraction $\phi=7.5\%$ in (c) and (d), and the distribution of coordination number $z$ in (e) and (f). The distributions here are obtained by averaging over $5$ independently generated structures and the error bars are smaller than the bar thickness. The persistence length $l_p$ and contour length $l_C$ of the strands for the above two angles $\bar{\theta}$ in (g). Here the data refer to $k_{B}T/\varepsilon = 10^{-2}$.}
\label{changeAngle}
\end{figure}
%%%%%%%%%%%%%%%%%%%%%%%%%%%%%%%%%%%%%%%%%%%%%%%%%%%%%%%%%%%%%%%%%%%%%%
Fig.~\ref{changeAngle} (top) shows the corresponding contour plots of the potential energy $u_\mathrm{total}$, indicating how shifting the value of the angle $\bar{\theta}$ towards higher values decreases the region where bonding to the central particles can occur. We expect therefore that gels are less likely to form in this case, since 3-coordinated structures are limited. The simulation snapshots and distributions of $z$ in Fig. \ref{changeAngle} (e,f) indeed show that the fraction of particles with $z=3$ is greatly reduced with increasing $\overline{\theta}$. Particles prevalently aggregate into strands that are one particle diameter thick and have much weaker tendency to branch, and hence of forming a gel network.
The persistence length and the average contour length $l_C$ of these particle strands are compared in Fig. \ref{changeAngle}(g). The structures obtained at $\bar{\theta}=75^\circ$ are softer since $l_C>l_p$ in contrast to one obtained for $\bar{\theta}=65^\circ$ where the two length-scales are comparable, feature similar to semiflexible networks \cite{Storm2005,Broedersz2014,Licup2015}. Finally, we can also vary the parameter $w$, which sets the width of peak of $u_3$ (see Fig. \ref{potentials} (b)). Decreasing $w$ makes the peak narrower, resulting in a sharper transition from attraction to repulsion when we consider the effective potential represented in the contour plot of Fig. \ref{fig2} (b). For smaller $w$, the region in blue (where the attraction dominates) in the contour plots becomes wider, favoring locally more compact microstructures. Hence, $w$ can be used together with $B$, $\bar{\theta}$ and the cooling rate $\Gamma$, to tune the gel microstructure.  

To summarize, the discussion in this section elucidates how varying the parameters in the model and in particular those entering the three-body term, changes the potential energy surfaces that drive the particles self-assembly and provides constraints to the local structures, as indicated by the contour plots for $u_{total}$. We can think of these local structures as the building blocks of the mesoscale gel network, hence the changes of the potential energy surfaces have also implications for the gel self-assembly as in fact demonstrated by the snapshots.

Ultimately, the gel structures depend of course on the kinetics of the self-assembly and on the gelatiion protocol. We can vary the gelation protocol, for example, through changing the cooling rate or other aspects of the procedure described in section \ref{preparation}. However, 
understanding how specific microscopic interactions can modify the local structures and promote distinct gel characteristics such as local coordination, mesoscale aggregates or specific ranges of pore size distributions can help systematically tune the gel structure. To build the link between the microscopic interactions and the emerging elasticity of the gel network, a first step is to consider what are the mechanical properties of the local structures which are eventually embedded in the gel. Therefore, in the remainder of the paper, we study the elasticity of local elementary structures that tend to form for the set of potential parameters used in Figs. \ref{potentials}, \ref{fig2} and \ref{fig3}(a,d,g). We analyze how the elasticity of these local structures can be estimated, and discuss how they can contribute to the emerging elastic properties of the gel network.  

\section{Elasticity of local structural elements and their contribution to the gel network elasticity}\label{elasticity}

We now consider the specific set of parameter values $B = 67.27$, $\overline{\theta}=65^{\circ}$ and $w = 0.30$. This set has been used in various other works to produce gels with open network structures as demonstrated above, also comparing dynamics and mechanical response with experiments \cite{Colombo:2014SM,Colombo:2014JOR,Bouzid2017,Bouzid:2018LAN,BouzidJOR2018,Feng:2018PNAS,Vereroudakis2020}. In these gels, the building units are \textit{strands} that are one particle diameter thick (particles in strands have coordination number $z=2$) and are connected through \textit{branching points} (that have coordination number $z=3$). Hence, here we examine how the elasticity of the gel networks for this choice of the parameters may result from the elastic properties of these local structural elements. 

Previous studies have shown that these structural elements can contribute to both stretching and bending terms in the overall linear and non-linear response of the gel network \cite{Colombo:2014JOR,Bouzid:2018LAN}.  Similar to semiflexible filaments in biopolymer networks, stretching and bending contributions to elastic stresses are comparable in the linear regime, whereas stretching becomes prevalent in the non-linear regime \cite{Broedersz2014}. 

\subsection{Stretching and bending moduli of particle strands}\label{modulus}
To estimate the stretching modulus of the gel strands, 
let us consider two bonded particles separated by a distance corresponding to the minimum of the attractive well in $u_{2}$, ($r_{min} = (\frac{18}{16})^{1/2}d \approx 1.06d$) (see Fig.  \ref{potentials}(a)). To estimate the stretching force constant ($k_s$), we consider a stretch $dr$ and, within a harmonic approximation, compute the curvature 
%%%%%%%%%%%%%%%%%%%%%%%%%%%%%%%%%%%%%%%%%%%%%%%%%%%%%%%%%%%%%%%%%%%%%%
\begin{align}
  k_s &= \Big[\frac{d^2u_2(r)}{dr^2}\Big]_{r=r_{min}}
      =\frac{736\varepsilon}{d^2}\Big ( \frac{d}{r_{min}}\Big)^{18}.
\end{align}
%%%%%%%%%%%%%%%%%%%%%%%%%%%%%%%%%%%%%%%%%%%%%%%%%%%%%%%%%%%%%%%%%%%%%%
The total energy of a strand made of $N$ bonds and stretched from its original length $L_0 = Nr_{min}$ by $\delta L=Ndr$ is given by 
\begin{align}
    U_{stretch}&=\frac{1}{2} k_sr_{min} (Nr_{min}) (\delta L/L_{0})^2\\
    &= \frac12 k_sr_{min}\int ds (\delta L/L_{0})^2.
\end{align}
By comparing the stretching energy typically used for a semiflexible polymer strand \cite{Broedersz2014,Licup2015}, the stretching modulus of a strand in our microscopic model is $\sigma_S=k_sr_{min}=\frac{736\varepsilon}{d}( d/r_{min})^{17}$, giving us $\sigma_S \approx 273 \varepsilon/d$ for the set of microscopic parameters considered here. The stretching mode becomes dominant at larger deformations when the bonds can no longer reorient and need to be stretched out to accomodate the deformation, contributing to the nonlinear elasticity.

Bending one of the particle strands also costs energy. 
For a strand composed of $N_p$ particle bonds, the total bending energy $U_\mathrm{st}$ can be computed as :
 %%%%%%%%%%%%%%%%%%%%%%%%%%%%%%%%%%%%%%%%%%%%%%%%%%%%%%%%%%%%%%%%%%%%%%
 \begin{align}
 U_\mathrm{st}&=\sum_{i=1}^{N_p-1} \frac{k_{\theta, \mathrm{st}}}{2}(\theta_{i+1}-\theta_i)^2\\
 &=\sum_{i=1}^{N_p-1} \frac{k_{\theta, \mathrm{st}}}{2}\bigg(\frac{\partial \hat{t}}{\partial s}\bigg)_i^2(r_{min})^2=\int ds\frac{k_{\theta, \mathrm{st}} r_{min}}{2}\bigg(\frac{\partial \hat{t}}{\partial s}\bigg)^2
 \end{align}
 %%%%%%%%%%%%%%%%%%%%%%%%%%%%%%%%%%%%%%%%%%%%%%%%%%%%%%%%%%%%%%%%%%%%%%
where $\hat{t}=\partial \vec{r}/\partial s$ is the unit vector tangent to the strand and $k_{\theta, \mathrm{st}}$ is the bending force constant of the strand. From this expression, the bending modulus of our strand is $\kappa_\mathrm{st}=k_{\theta, \mathrm{st}} r_{min}$ \cite{Broedersz2014,Licup2015}.

For each segment of three particles, if the bond angle is changed by $\delta \theta = \pi-\theta$ from its equilibrium configuration to a new configuration (see Fig. \ref{fig2} (a)), the energy of its new configuration is:
%%%%%%%%%%%%%%%%%%%%%%%%%%%%%%%%%%%%%%%%%%%%%%%%%%%%%%%%%%%%%%%%%%%%%%
\begin{align}\label{uchain}
%U_{3,\mathrm{branching}} & =
U_\mathrm{st} & =
\begin{aligned}[t]
    & 2u_2(r)+u_2(2r\cos \left (\delta \theta /2\right)+u_3(r,r,\pi-\delta \theta)\\
    &+2u_3(r,2r\cos \left (\delta \theta /2\right),\delta \theta/2).\\
\end{aligned}
\end{align}
%%%%%%%%%%%%%%%%%%%%%%%%%%%%%%%%%%%%%%%%%%%%%%%%%%%%%%%%%%%%%%%%%%%%%%. 
The curvature of this function provides the force constant $k_{\theta,\mathrm{st}}$ in response to the bending of a three particle segment. 

 %%%%%%%%%%%%%%%%%%%%%%%%%%%%%%%%%%%%%%%%%%%%%%%%%%%%%%%%%%%%%%%%%%%%%%
\begin{figure}[h]
\centering
\includegraphics[scale=0.11]{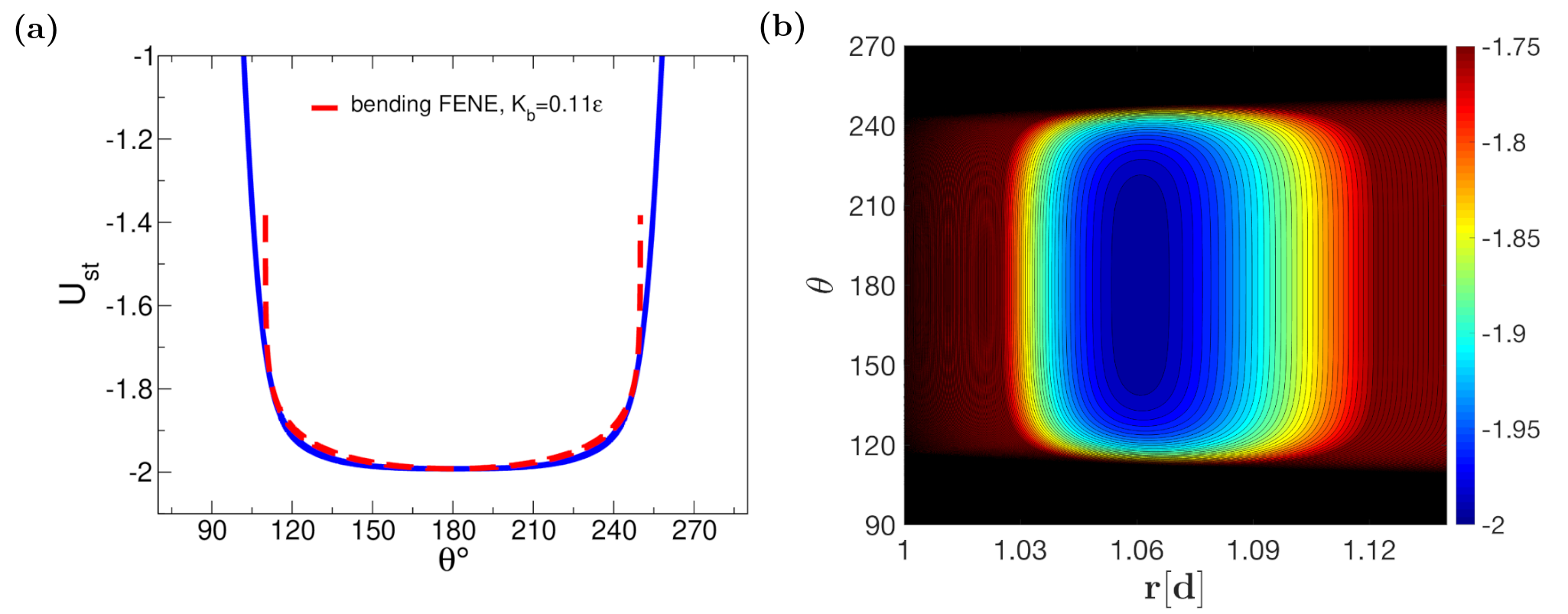}
\caption{(a) The potential energy profile for a 3-particle strand is shown in blue line and the fit of the bending FENE potential in red dashed line (b) The contour plot of the same energy $U_\mathrm{st}$ for different distances and angles with the color code representing the energy values.}
\label{fig3}
\end{figure}
%%%%%%%%%%%%%%%%%%%%%%%%%%%%%%%%%%%%%%%%%%%%%%%%%%%%%%%%%%%%%%%%%%%%%%

The energy profile of $U_\mathrm{st}(\theta)$ is plotted in Fig. \ref{fig3} (a): it is relatively flat in the middle and non-harmonic. The curvature changes slowly close to the minimum ($\theta=180^\circ$) while it grows rapidly for angles approaching $\theta=140^\circ$. This nonlinear dependence can be well approximated by a finitely extensible nonlinear elastic (FENE) type of potential of the form $U_\mathrm{st}=-\frac{1}{2}K_\mathrm{b} \Delta \theta_\mathrm{max}^2\ln{[1-(\frac{\theta-\theta_0}{\Delta \theta_\mathrm{max}})^2]}$, typically used for semiflexible polymers \cite{Kroger2005}. The fit shown in the plot (dashed line) corresponds to the parameters $\theta_0=180^\circ, \Delta \theta_\mathrm{max} \sim 68^\circ$ and $K_\mathrm{b} = 0.11\varepsilon$, from which  we obtain the microscopic bending force constant $k_{\theta,\mathrm{st}} \approx K_b \approx 0.11\varepsilon$ and the bending modulus $\kappa_{\mathrm{st}} \approx 0.12\varepsilon d$. 

 %%%%%%%%%%%%%%%%%%%%%%%%%%%%%%%%%%%%%%%%%%%%%%%%%%%%%%%%%%%%%%%%%%%%%%
\begin{figure}[htp]
\centering
\includegraphics[scale=0.2]{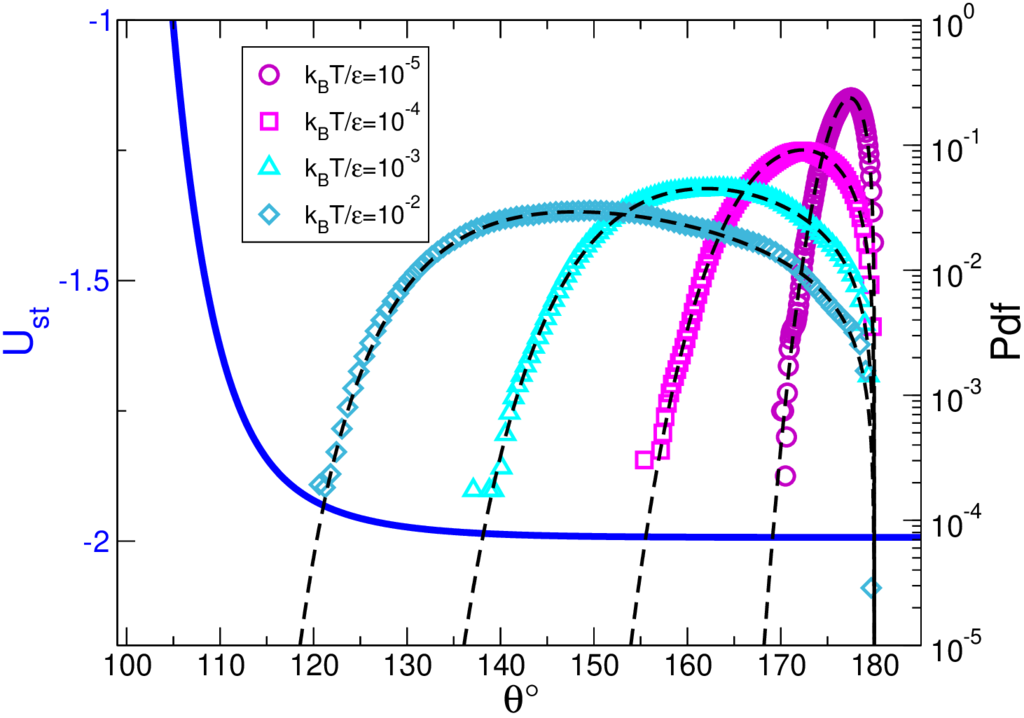}
\caption{The energy profile for $U_\mathrm{st}$ (left axis) in blue solid line. The bond angle distributions (right axis) for 3-particle strand for different thermal fluctuations. The dashed lines represent the prediction from Boltzmann probability distribution.}
\label{BondAnglesChangeTemp}
\end{figure}
%%%%%%%%%%%%%%%%%%%%%%%%%%%%%%%%%%%%%%%%%%%%%%%%%%%%%%%%%%%%%%%%%%%%%%
%%%%%%%%%%%%%%%%%%%%%%%%%%%%%%%%%%%%%%%%%%%%%%%%%%%%%%%%%%%%%%%%%%%%%%
\begin{figure*}[htp]
\centering
\includegraphics[scale=0.17]{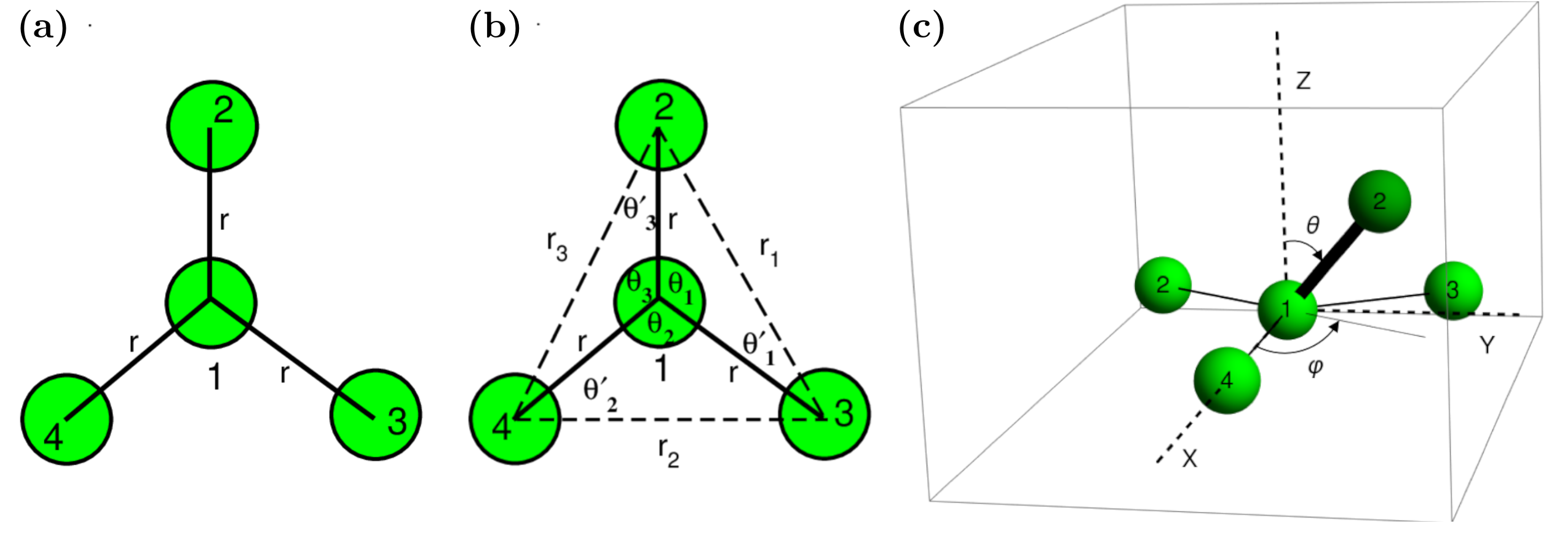}
\caption{(a) Planar branching point: The particle labeled \textbf{1} at the center is a branching point, and it is bonded to three other particles \textbf{2}, \textbf{3} and \textbf{4} with all at a distance $r$. (b) Representation of the distances and angles for the calculation of energy. (c) 3-D representation of planar and non planar branching points: The four light green particles lying on XY-plane form a planar configuration. The three green particles and a dark green particle form a non-planar configuration. The position of the dark green particle is represented in polar coordinates ($r,\theta,\varphi$). }
\label{SchematicBranchingPoint}
\end{figure*}
%%%%%%%%%%%%%%%%%%%%%%%%%%%%%%%%%%%%%%%%%%%%%%%%%%%%%%%%%%%%%%%%%%%%%%
Fig. \ref{fig3} (b) shows the contour plot of $U_\mathrm{st}$ for a strand made of three particles as in Fig. \ref{fig2} (a), varying the distance and the bond angle of particle \textbf{3}, with the color providing information on the energy values. The plot shows that the energy profile is symmetric along the direction $\theta$ (as seen in Fig. \ref{fig3} (a)) but is instead asymmetric along $r$, indicating that the curvature of $U_\mathrm{st}$ along $\theta$, and hence the bending modulus, varies with the particle separation, as a result of the strong overall dependence of the microscopic interactions on interparticle separation. We have tested these calculations using MD simulations of three particle strand segments to compute the distribution of bond angles obtained with different amount of thermal fluctuations. The results are shown in Fig. \ref{BondAnglesChangeTemp} for different ratios $k_BT/\varepsilon$. With increasing thermal energy, the distributions become wider and extend to smaller angles, but in all cases follow a Boltzmann probability distribution for bending angles of the form $p(\theta) d\theta \sim \sin \theta \exp[-U_\mathrm{st}(\theta)/k_BT] d\theta$.

\subsection{Stretching and bending moduli of branching points}\label{modulusbr}
\begin{figure*}[htp]
\centering
\includegraphics[scale=0.085]{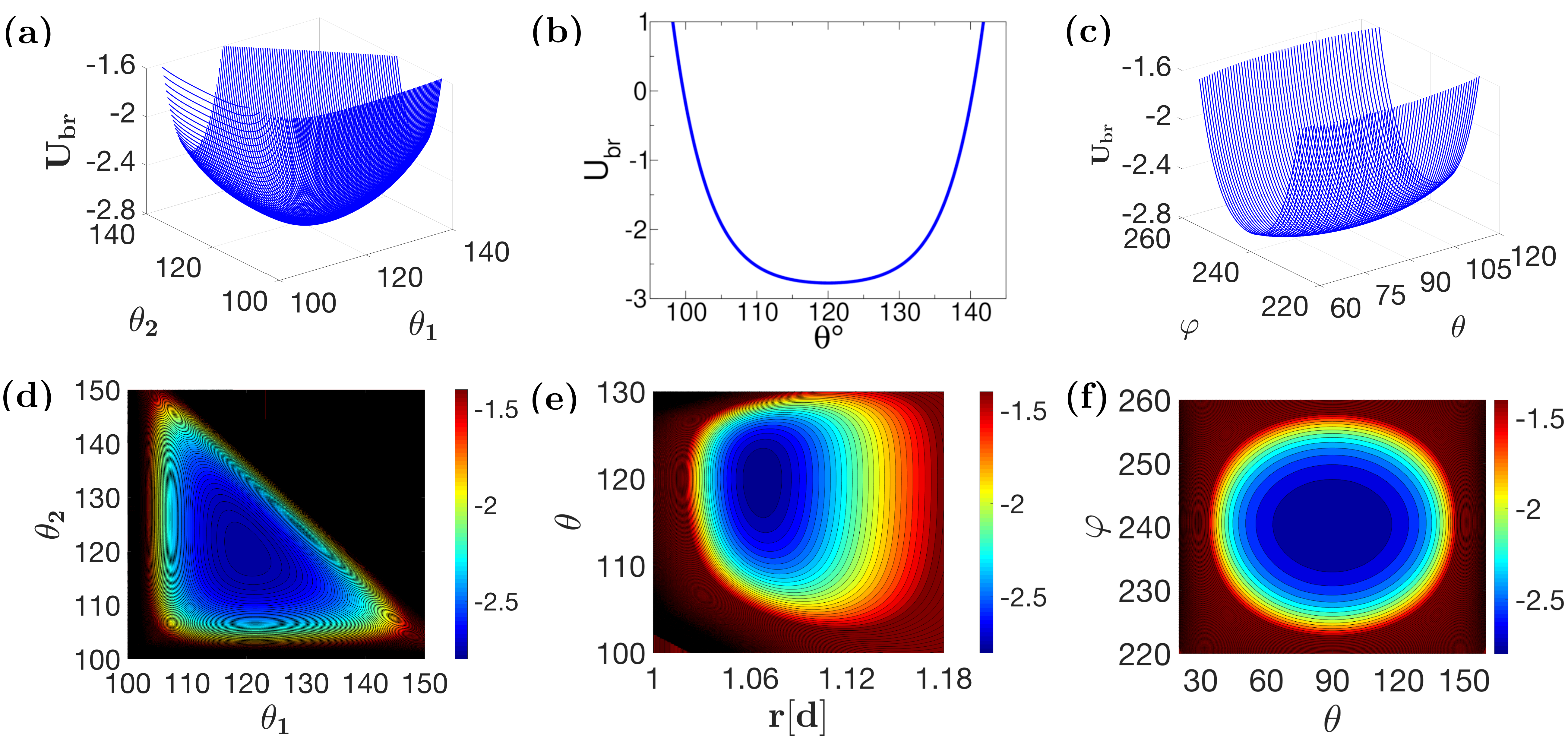}
\caption{Energy profiles for a branching point: (a) The total potential energy $U_\mathrm{branching}$ as a function of two angles $\theta_1$ and $\theta_2$ in a planar branching point, with all the particle distances are fixed at $r=r_{min}$. (b) $U_\mathrm{br}$ as a function of angle $\theta$ along the diagonal $\theta_1=\theta_2=120+\theta$. (c) The total potential energy $U_\mathrm{br}$ in a non planar branching point, with all the particle distances fixed at $r=r_{min}$. (d) The total potential energy contour for varying angle $\theta$ and distance $r$, for the planar case.
(e) The contour of total potential energy $U_\mathrm{br}$ as a function of two angles $\theta_1$ and $\theta_2$ in the planar case. (f) The contour of total potential energy $U_\mathrm{br}$ as a function of $\theta$ and $\varphi$ in a non planar branching point (with all the particle distances fixed at $r=r_{min}$).}
\label{EnergyDiagramBranchingPoint}
\end{figure*}

In the case of a branching point, a particle is bonded to three other particles (with roughly the same bond lengths $r$) as represented in Fig.~\ref{SchematicBranchingPoint}. Planar branching configurations are shown in Fig \ref{SchematicBranchingPoint} (a) and (b). The case of a non-planar branching point is illustrated in Fig. \ref{SchematicBranchingPoint} (c). Here four particles shown as light green spheres initially form a planar configuration in the XY plane. The configuration becomes non-planar when the particle labeled \textbf{2} moves to the position represented by the dark green sphere and the thicker bond. This out-of-plane position can be expressed in terms of polar coordinates ($r,\theta,\varphi$), where $r$ is the radial distance, $\theta$ the polar angle and $\varphi$ the azimuthal angle. The potential energy of a branching configuration can be expressed in terms of distances and angles labeled in Fig. \ref{SchematicBranchingPoint} (b) as a sum of $u_2$ and $u_3$ terms as:
$U_\mathrm{br} = U_{2,\mathrm{br}}+U_{3,\mathrm{br}}$. The total $u_2$ potential is computed for all interaction pairs:
\begin{equation}\label{u2branching}
U_{2,\mathrm{br}}= 3u_2(r)+u_2(r_1)+u_2(r_2)+u_2(r_3)  
\end{equation}
%%%%%%%%%%%%%%%%%%%%%%%%%%%%%%%%%%%%%%%%%%%%%%%%%%%%%%%%%%%%%%%%%%%%%%
with the following distances 
%%%%%%%%%%%%%%%%%%%%%%%%%%%%%%%%%%%%%%%%%%%%%%%%%%%%%%%%%%%%%%%%%%%%%%
\begin{equation}\label{parabranching1}
r_1 = 2r \sin \left(\frac{\theta_1}{2} \right), \;\; r_2 = 2r \sin \left(\frac{\theta _2}{2}\right),\;\; r_3 = 2r \sin \left(\frac{\theta _3}{2}\right) 
\end{equation}
%%%%%%%%%%%%%%%%%%%%%%%%%%%%%%%%%%%%%%%%%%%%%%%%%%%%%%%%%%%%%%%%%%%%%%
Let us consider first the case of a planar branching point as in Fig. \ref{SchematicBranchingPoint} (b), so that $\theta _3 = 2 \pi - (\theta _1 + \theta _2)$. To compute $U_{3,\mathrm{br}}$, we need to consider the total $u_3$ potential for a given triangle by taking the $u_3$ terms from Eq. \ref{utotal1}. In such case, the total $u_3$ potential is the sum of contributions from three small triangles formed by the particles \textbf{1}-\textbf{2}-\textbf{3}, \textbf{1}-\textbf{3}-\textbf{4} and \textbf{1}-\textbf{2}-\textbf{4}, and a larger triangle formed by particles \textbf{2}-\textbf{3}-\textbf{4} as follows: 
%%%%%%%%%%%%%%%%%%%%%%%%%%%%%%%%%%%%%%%%%%%%%%%%%%%%%%%%%%%%%%%%%%%%%%
\begin{align}\label{u3branching}
U_{3,\mathrm{br}} & =
\begin{aligned}[t]
& u_3(r,r,\theta_1)+u_3(r,r,\theta_2)+u_3(r,r,\theta_3)\\
&+2u_3(r,r_1,\theta^\prime_1)+2u_3(r,r_2,\theta^\prime_2)+2u_3(r,r_3,\theta^\prime_3)\\
&+ u_3(r_1,r_2,\theta^\prime_1+\theta^\prime_2)+u_3(r_1,r_3,\theta^\prime_1+\theta^\prime_3)\\
&+u_3(r_2,r_3,\theta '_2+\theta^\prime_3)
\end{aligned}
\end{align}
%%%%%%%%%%%%%%%%%%%%%%%%%%%%%%%%%%%%%%%%%%%%%%%%%%%%%%%%%%%%%%%%%%%%%%
with distances and angles for a planar branching point:
%%%%%%%%%%%%%%%%%%%%%%%%%%%%%%%%%%%%%%%%%%%%%%%%%%%%%%%%%%%%%%%%%%%%%%
\begin{equation}\label{parabranching2}
r_i = 2r\sin \left(\frac{\theta_i}{2}\right), \;\;\; \theta '_i = (\pi - \theta_i)/2, \; i=1,2,3. 
\end{equation} 
%%%%%%%%%%%%%%%%%%%%%%%%%%%%%%%%%%%%%%%%%%%%%%%%%%%%%%%%%%%%%%%%%%%%%%
Using Eq. \ref{u2branching} - \ref{parabranching2}, the resulting potential $U_\mathrm{br}$ is therefore only a function of three variables $r$, $\theta _1$ and $\theta _2$. It is computed as a function of the angles $\theta _1$ and $\theta _2$ by fixing the distance $r=r_{min}$ in order to generate a 3-D plot in Fig. \ref{EnergyDiagramBranchingPoint} (a) and a contour plot, with the color code representing the potential energy values, shown in Fig. \ref{EnergyDiagramBranchingPoint} (d). These plots indicate that the minimum of the potential energy depends on the combination of $\theta _1$ and $\theta _2$ with the global minimum at $\theta=\theta _1=\theta _2=120^\circ$. The energy profile along the diagonal $\theta_1=\theta_2=120+\theta$ is shown in Fig. \ref{EnergyDiagramBranchingPoint} (b), it is approximately harmonic and the microscopic bending force constant can be estimated from the curvature at the minimum of the well. The energy profile is also dictated by the distance $r$ and varies asymmetrically as shown in Fig. \ref{EnergyDiagramBranchingPoint} (e).
%%%%%%%%%%%%%%%%%%%%%%%%%%%%%%%%%%%%%%%%%%%%%%%%%%%%%%%%%%%%%%%%%%%%%%
%%%%%%%%%%%%%%%%%%%%%%%%%%%%%%%%%%%%%%%%%%%%%%%%%%%%%%%%%%%%%%%%%%%%%%

We estimate the effective response  of a planar branching point to bending by assuming that the two angular degrees of freedom in a planar branch respond in a manner similar to two springs in series, so that their combination is dominated by the softest of the two. In this approximation, the effective bending force constant is given by the sum of the inverse of the two spring constants along the principal directions of the Hessian $H_{ij}=\partial_{\theta_i}\partial_{\theta_j} U_\mathrm{br}(\theta_1,\theta_2)$. That is, we calculate $k_{\theta,\mathrm{br}}^{-1} = 1/H_{11} + 1/H_{22}$ using a discretization along two directions $\theta_1$ and $\theta_2$ in Fig. \ref{EnergyDiagramBranchingPoint} (a) as follows: 
%%%%%%%%%%%%%%%%%%%%%%%%%%%%%%%%%%%%%%%%%%%%%%%%%%%%%%%%%%%%%%%%%%%%%%
\begin{equation} \label{surfacecurvature}
k_{\theta,\mathrm{br}}^{-1} =\hspace{2.9in}
\end{equation}
\begin{align}
\begin{aligned}[t]
& \frac{\delta \theta _1 ^2}{U_{\mathrm{br}} (r,\theta _1+ \delta \theta _1, \theta _2)+U_{\mathrm{br}} (r, \theta _1 - \delta \theta _1, \theta _2)-2U_{\mathrm{br}} (r, \theta _1, \theta _2)}\\
& + \frac{\delta \theta _2 ^2} {U_{\mathrm{br}} (r, \theta _1 , \theta _2+\delta \theta _2)+U_{\mathrm{br}} (r, \theta _1, \theta _2 - \delta \theta _2)-2U_{\mathrm{br}} (r, \theta _1, \theta _2)} \\
\end{aligned}
\end{align}
%%%%%%%%%%%%%%%%%%%%%%%%%%%%%%%%%%%%%%%%%%%%%%%%%%%%%%%%%%%%%%%%%%%%%%
with $\delta \theta _1=0.01^\circ$  and $\delta \theta _2=0.01^\circ$ the spacing between angles on the calculation grid. The curvature is computed on the grid from angles $\theta _1 = \theta _2 = 115^{\circ}$ to $\theta _1 = \theta _2 = 125^{\circ}$ in Fig. \ref{EnergyDiagramBranchingPoint} (a), and its value is determined to be $k_{\theta, \mathrm{br}} \approx 5.44 \varepsilon$. The bending modulus is then given by $\kappa_\mathrm{br}=k_{\theta,\mathrm{br}}r_{min} \approx 5.77 \varepsilon d$, a value approximately $54$ times higher than the bending modulus for a strand. These results suggest that increasing the amount of branching points should dramatically increase the gel modulus, as indeed found in simulations \cite{Bouzid:2018LAN}. 
 %%%%%%%%%%%%%%%%%%%%%%%%%%%%%%%%%%%%%%%%%%%%%%%%%%%%%%%%%%%%%%%%%%%%%%
\begin{figure*}[htp]
\centering
\includegraphics[scale=0.19]{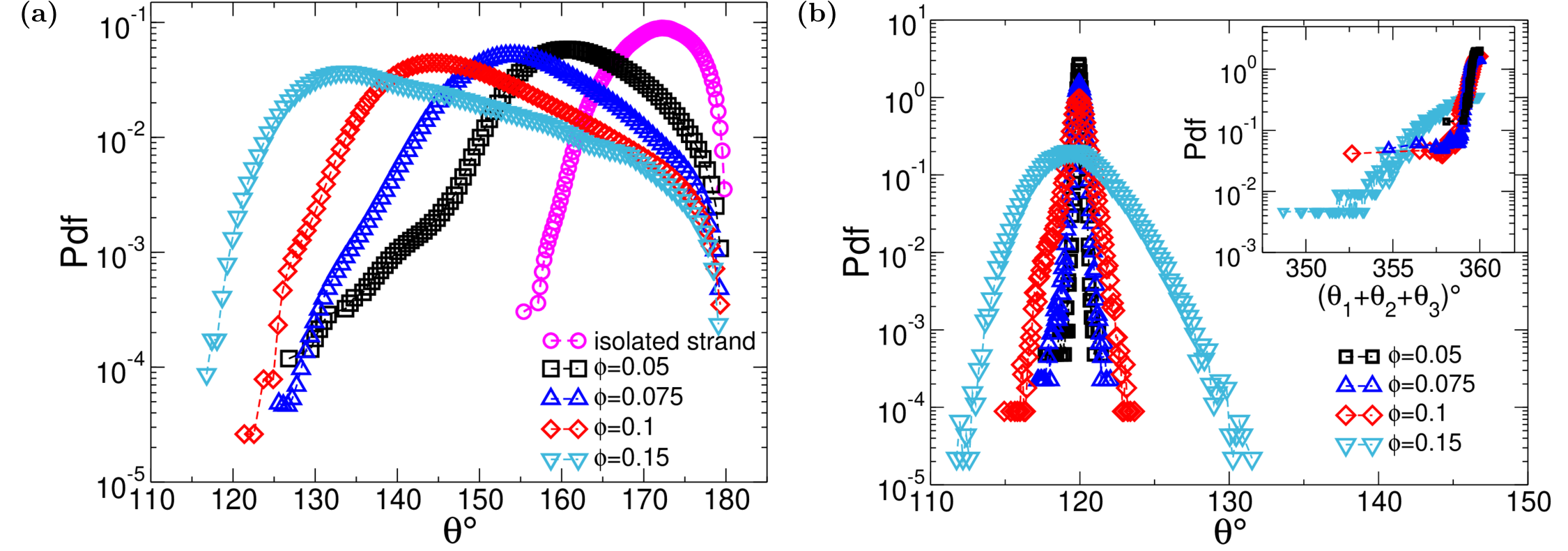}
\caption{(a) The distribution of bond angles for an isolated strand and for strands in networks at different volume fractions $\phi$. (b) Main: The distributions of bond angles that correspond to the branching points. Inset: The distribution of sum of three angles at a single branching point where the sum equal to 360$^\circ$ corresponds to a planar branching point.}
\label{BondAnglesSimulation}
\end{figure*}
%%%%%%%%%%%%%%%%%%%%%%%%%%%%%%%%%%%%%%%%%%%%%%%%%%%%%%%%%%%%%%%%%%%%%%
For non-planar branching configurations, particles are no longer restricted to a single plane and the angles are not bound by the constraint $\theta_1+\theta_2+\theta_3=360^\circ$. We can determine the angle between each of the neighboring bonds in Fig. \ref{SchematicBranchingPoint} (c) in terms of the angles $\theta$ and $\varphi$. Using the same convention for angles and distances as in the planar case, the central angle of a triangle formed by particles \textbf{3-1-4} is $\theta_2$. Considering the center of the central particle \textbf{1} in the branching point at the origin of the Cartesian coordinate system C$_1(0,0,0)$, we can write the coordinates of the centers of the other particles in Fig. \ref{SchematicBranchingPoint} (c) as C$_4(r,0,0)$, C$_3(r\cos\theta_2,r\sin\theta_2,0)$ and C$_2(r\sin\theta\cos\varphi,r\sin\theta\sin\varphi,r\cos\theta)$. We can now express the central angle of a triangle formed by the particles \textbf{2-1-3} in terms of ($\theta$, $\varphi$) as $\theta_1=\cos^{-1}{(\sin\theta \cos(\theta_2-\varphi))}$. Similarly for the triangle \textbf{2-1-3}, we obtain $\theta_3=\cos^{-1}{(\sin\theta \cos\varphi)}$. This allows us to determine all the angles and distances in each of the triangles. Finally, the total energy of a non-planar branching point in Eqs. \ref{u2branching} and \ref{u3branching} is a function of only $\theta_2,\theta$ and $\varphi$. In Fig.~\ref{EnergyDiagramBranchingPoint} (c), we plot the total potential energy $U_\mathrm{br}$ as a function of $\theta$ and $\varphi$ for $\theta_2=120^\circ$. The potential is asymmetric along the $\theta$ and $\varphi$ directions and the minimum of the potential is at $\theta = 90^{\circ}$ and $\varphi=240^\circ$ which corresponds to a planar branching point (see Fig. \ref{EnergyDiagramBranchingPoint} (f)). We compute the energy surface curvature from Fig. \ref{EnergyDiagramBranchingPoint} (c) as done for the planar branching point and find the curvature along $\theta$ to be $k^{np}_{\theta} \approx 0.66 \varepsilon$, and along $\varphi$ to be $k^{np}_{\varphi} \approx 10.88 \varepsilon$. We combine again these two curvatures as springs in series to obtain $k^{np}_\mathrm{br} \approx 0.62 \varepsilon$. All bending constants estimated for the different types of local structures are summarized in Table \ref{table2}.
\begin{table}[h!]
\centering
\caption{{Estimated bending constants for different local structures.}}
\label{table2}
\begin{tabular}{|c|c|c|}
\hline
{Local structure} & {Symbol} & {Estimated value}\\
\hline
    {Strand} & {$k_{\theta, \mathrm{st}}$} & {0.11 $\varepsilon$}\\
\hline
  {Planar branching point} & {$k_{\theta, \mathrm{br}}$} & {5.44 $\varepsilon$}\\
\hline
  {Non-planar branching point} & {$k^{np}_{\theta, \mathrm{br}}$} & {0.62 $\varepsilon$}\\
\hline
\end{tabular}
\end{table}

 \subsection{Gel elasticity}
 \label{gelmodulus}
 We now complement the insight obtained from the energy profiles for the different local structures (strands and branching points) with the one obtained from MD simulations of gel networks for the choice of parameters of interest here, which correspond to a gel made of one particle thick strands connected by branching points.  
 
All calculations and estimates made in sections \ref{modulus} and \ref{modulusbr} correspond to the elastic properties of isolated strands or branching points.
 In the simulations, we can investigate instead how the bond angle distributions change for strands (coordination number $z=2$) and branching points ($z=3$) that are embedded in gel networks where they experience topological constraints due to the fact that they are connected to each other and coupled across the network. In Fig. \ref{BondAnglesSimulation} (a), we see that the bond angle distribution for the strands changes significantly with the solid volume fraction $\phi$ of the gel network, with a peak that shifts to smaller angles with increasing $\phi$.
 Hence, when connected in the network, the bending stresses experienced by the strands can be very different. In Fig.~\ref{BondAnglesSimulation} (b) we show the distributions of bond angles obtained from the branching points of the gel network in the MD simulations. The distributions of angles over different branching points is obtained in terms of the sum of all three angles formed at the central particle, shown in the inset of Fig. \ref{BondAnglesSimulation} (b). The sum equal to $360^\circ$ corresponds to planar configurations. We can see that the fraction of non-planar configurations increases with increasing the gel volume fraction and has stronger tails for sum of angles smaller than $360^\circ$ for $\phi=0.15$. These distributions are always strongly peaked at $\theta = 120^\circ$, indicating that most branching points configurations can be well captured by the simple planar approximation at low enough volume fractions. The tails of these distributions, however, clearly widen for gels at higher particle volume fractions, which also correspond to an increase of the density of branching points \cite{Bouzid:2018LAN}. The data at the highest $\phi$ considered is clearly distinct. The changes with $\phi$ in the two sets of distributions plotted in Fig. \ref{BondAnglesSimulation} indicate that increasing the particle volume fractions and density of branching points introduces stronger constraints on the angles of the local structures that compose the gels. The constraints are mainly topological in nature, i.e. they emerge from the topology and connectivity of the gel network rather than from the direct steric or bonding interactions between the particles in each elements. These findings suggest that, for both strands and branching points,
 displacements and fluctuations must be increasingly hindered and correlated upon increasing the gel volume fraction. Such effects should have an impact on the gel elasticity, suggesting that the network elasticity can not be anymore obtained just from that of the isolated structural elements. 

To estimate the different contributions of the bending elasticity of the isolated structural elements to the total elasticity of the network, we can consider, in a first approximation, that the contribution of branching points is given by their elastic energy per unit volume, and is obtained as a discrete sum over the different angles in both the planar branching and non-planar branching points. Therefore, we sum over the angle distributions (Fig. \ref{BondAnglesSimulation} (b)) to obtain
\begin{equation}
K_\mathrm{br} = \frac{1}{2V}\sum_\theta [n_\theta k_{\theta, \mathrm{br}}(\theta-\theta_0)^2 + n^{np}_\theta k^{np}_{ \mathrm{br}}(\theta-\theta_0)^2]
\end{equation}
where $k_{\theta, \textrm{br}}=5.44 \varepsilon$ is the bending force constant for a planar branching point and $k^{np}_{ \mathrm{br}}=0.62 \varepsilon$ for the non-planar branching points (see Table \ref{table2}), $\theta_0$ is the position of the peak in the distribution, $n_\theta$ and $n^{np}_\theta$ are the counts for each angle $\theta$ respectively in a planar and non-planar case and $V$ is the volume of the simulation box.
We also estimate the bending elasticity contribution from the strands using the bond angle distributions in Fig. \ref{BondAnglesSimulation} (a). For each bond angle $\theta$, if there are $n_\theta$ three-particle segments in the strands from which we collect the distribution, the elastic energy density of the strands is obtained by summing over the distribution as
\begin{equation}
 K_\mathrm{st} = \frac{1}{2V}\sum_\theta n_\theta k_{\theta, \mathrm{st}}(\theta-\theta_0)^2
    \end{equation}
where $k_{\theta,\mathrm{st}}=0.11\varepsilon$ is the bending force constant for three-particle chains obtained from the energy profile in Fig. \ref{fig3} (a), and $\theta_0=180^\circ$ is the angle corresponding to the minimum of the same energy profile. Finally, we also consider the total elastic contribution from strands and branching points in the limiting cases of the two springs of moduli $K_\mathrm{st}$ and $K_\mathrm{br}$ either in parallel giving $K^p_{\mathrm{total}} = K_\mathrm{st} + K_\mathrm{br}$ or in series with $1/K^s_{\mathrm{total}} = 1/K_\mathrm{st} + 1/K_\mathrm{br}$. 
%%%%%%%%%%%%%%%%%%%%%%%%%%%%%%%%%%%%%%%%%%%%%%%%%%%%%%%%%%%%%%%%%%%%%%
\begin{figure}[htp]
\centering
\includegraphics[scale=0.22]{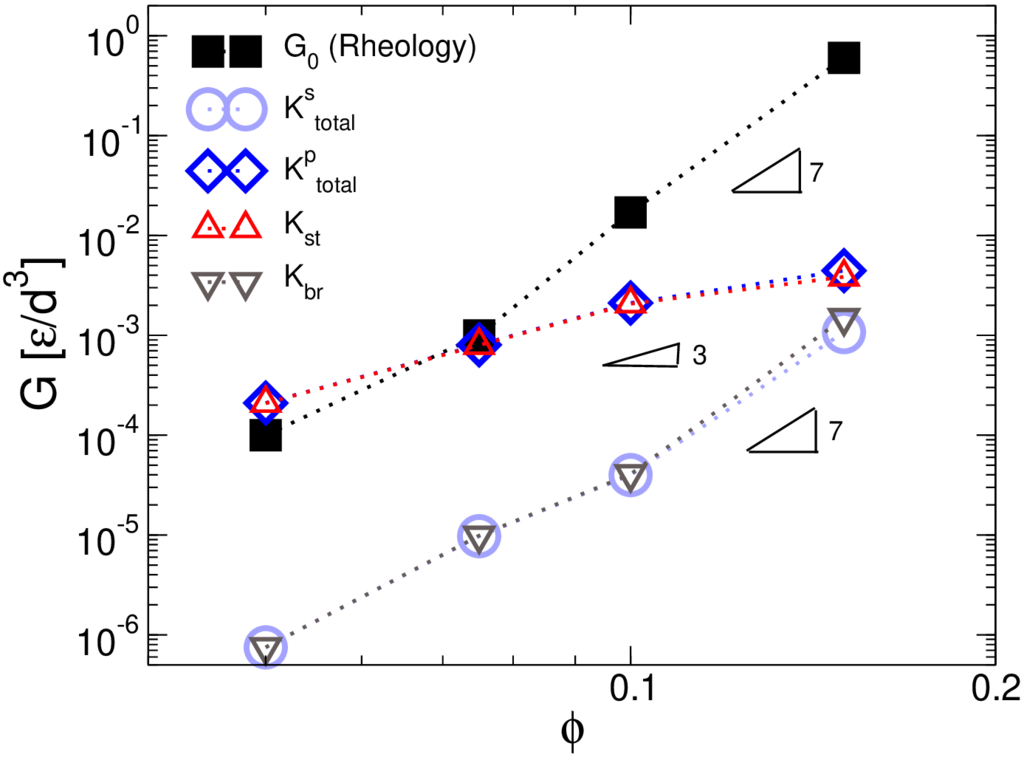}
\caption{Shear modulus of the gel networks measured from linear oscillatory rheology as a function of the volume fraction $\phi$ (squares). The value reported here corresponds to the low frequency plateau obtained from the viscoelastic spectrum measured on $3$ independently generated samples. The triangles symbols correspond to the contribution of respectively strands (up) and branching points (down), estimated using their bending moduli and the distribution of angles obtained from the simulations. The circle and diamond symbols correspond to the total contribution from the combination of strand and branching point contributions as two springs in series and parallel respectively.}
\label{EstimateNetworkElasticity}
\end{figure}
%%%%%%%%%%%%%%%%%%%%%%%%%%%%%%%%%%%%%%%%%%%%%%%%%%%%%%%%%%%%%%%%%%%%%%

The comparison between the plateau modulus obtained from the linear viscoelastic response of the gels \cite{Bouzid:2018LAN} and the estimates of the elastic energy density obtained from strands and branching point is shown in Fig. \ref{EstimateNetworkElasticity}, as a function of the solid volume fraction in the gels. The plateau moduli vary in the range $G_0 \sim [10^{-4}-1] \varepsilon/d^3$, which, for colloidal particles with $d\sim 100$ nm and $\varepsilon \sim 10k_BT$ would correspond to $G_0 \sim [5.10^{-3}-50]$ Pa, in relatively good agreement with typical values in experiments \cite{laurati2009,Koumakis2011,Helgeson2014,Whitaker2019,Vereroudakis2020,Gisler1999}. We can see that the total estimated elastic energy density is largest when the contributions from strands and branching points are combined as two springs in parallel $K^p_{\mathrm{total}}$ and smallest when they are connected in series $K^s_{\mathrm{total}}$. For all volume fractions, the main contribution to the modulus seems to be the one of the gel strands, which can be understood by considering that the strands are the majority component of the gel and that their bending costs less energy than for the branching points, hence dominating the linear response of the system. At low enough volume fractions, the bending energy of the strands seems to be enough to make up for the macroscopic behavior of the material. Upon increasing the volume fraction, however, it becomes clear that, while the contribution of the branching points is still lower in magnitude, they play a leading role: the dependence of the elastic modulus of the gels on the volume fraction is much stronger than the one of the elastic energy density due to the strands bending and seems to follow the same dependence of the bending energy density of the branching points. While the bending of the strands may be the main source of elasticity, the contribution of the branching points increases dramatically because, with increasing the volume fraction, not only the amount of branching point increases but also the connections between them (and hence their coupling) become important. 
The effect of feedback and coupling of the local structures is obviously not contained in the estimates of the elastic properties of the isolated element contributions. The comparison in Fig. \ref{EstimateNetworkElasticity} suggests that when we estimate the modulus by just summing up, as independent, the contributions of strands and branching points in the gel, we cannot account for the modulus actually measured in the viscoelastic tests. Only at low enough volume fraction are the two estimates close, as it is reasonable to expect. The same elastic contributions, obtained from {\it isolated} elastic elements can be mechanically combined in different ways in the gel network. For example, if we combine the different elastic contributions of the elastic elements ($K_\mathrm{st}$ and $K_\mathrm{br}$) as all springs in parallel or all in series, we obviously obtain very different values of the elastic modulus, either completely dominated by the strands or by the branching points stiffness (see Fig. \ref{EstimateNetworkElasticity}). This simple example indicates how the modulus of the network can depend strongly on the way the different elastic elements are combined through the network architecture. However, the specific combination corresponding to a certain network architecture is simply not known, nor there is an obvious way to predict it, because of the disorder and heterogeneities.

The results in Fig. \ref{EstimateNetworkElasticity} show how the single components we have identified in the gel networks are in fact not independent and they can only be approximated as such in the limit of very dilute and tenuous gels, i.e. at low enough $\phi$. With increasing $\phi$, not only there are more branching points, but also the way they are distributed and constrained by the network topology is different, as indeed indicated by the distribution of bond angles in Fig. \ref{BondAnglesSimulation} (a) and (b). Hence the missing contribution to the modulus must be the coupling between the different components through the network architecture, that can only be neglected when the gel is sufficiently dilute and its structure sufficiently tenuous. The data in Fig. \ref{EstimateNetworkElasticity} also suggest a power law dependence of the various contributions and of the gel elastic modulus on the particle volume fraction (with the caveat that our range of volume fractions here is relatively limited). Such dependence is quite common in colloidal gels although  exponents reported in experiments are often between $\sim 3\--4$ \cite{DelGadoBookChapter2016, Buscall1988, Shih1990, Russel1993, Piau1999,Trappe2000,Prasad2003} and usually associated with fractal structures. The gel structures considered here are not self-similar and the values we find are closer to those reported in \cite{Wyss2005,Yanez1996, Ramalrishnan2004, Mellema2002} corresponding to quite higher values of the exponent. 

\section{Conclusions}\label{conclusions}
Particle contacts in soft particulate gels can be dominated by surface roughness, sticky patches or other surface heterogeneities that modify the local energy profiles of the aggregated structures. The complex nature of these contacts can therefore play a role in the gel morphology and ultimately affect the gel mechanics, for example introducing an effective resistance to bending of the bonds between particles or of parts of the gel structure. To capture these features, we have proposed a class of effective interactions models that include, in additional to the usual short range attractive interaction term typical of gelling colloidal suspensions, a three-body term which depends on the angle between bonds departing from a central particle and introduces bending costs in the elastic energy of particle aggregates. These interactions are expressed in a mathematical form that is computationally convenient and allows for large scale simulations.  

The different gels formed in the numerical simulations have helped us elucidate how different model parameters control the formation of different types of particles aggregates, leading to a range of gel morphologies. Varying the model parameters, therefore, we can span from gels made of thin strands connected through branching points to gels featuring thick branches and large pores.

For gels made of semi-flexible strands connected by branching points, we have computed bending costs for both these types of elastic elements directly from the microscopic interactions. By comparing our analytical calculations with the gel plateau moduli obtained through linear viscoelastic tests, we gain new insight into how these distinct elastic elements contribute to the emerging gel elasticity. Our calculations and numerical simulations indicate that the mechanical coupling of strands and branching points across the network, which is determined by the network topology and connectivity, dominate the dependence of the gel modulus on gel density or particle volume fraction.

\section*{Acknowledgments}
MB and EDG would like to thank Stefano Aime, Daniel Blair, Mehdi Bouzid, Jasper Immink, Peter Schurtenberger and Jeff Urbach for insightful discussions. We acknowledge financial support from Georgetown University and National Science Foundation (NSF DMR-2026842). PDO thanks the Ives Foundation for support.

\section*{References}
\bibliographystyle{unsrt}
\bibliography{harvard}

\end{document}